# Structural Commonalities in Amorphous Elemental Materials

I. Rodríguez[1], R. M. Valladares[2*], A. Valladares[2], D. Hinojosa-Romero[2], F. B. Quiroga[1], S. Calderón-Alba[2], R. S. Vilchis-Peyret[2] and A. A. Valladares[1]

[1] Instituto de Investigaciones en Materiales, Universidad Nacional Autónoma de México, Apartado Postal 70-360, Ciudad Universitaria, CDMX, Ciudad de México 04510, México.
[2] Facultad de Ciencias, Universidad Nacional Autónoma de México, Apartado Postal 70-542, Ciudad Universitaria, CDMX, Ciudad de México 04510, México.
*Corresponding author: Renela M. Valladares, renelavalladares@gmail.com

## ABSTRACT

In recent times, the research community has explored diverse structures and novel fabrication methods for amorphous solids. This work investigates structural trends among different classes of amorphous materials to identify universal commonalities and fundamental differences. It is found that amorphous semiconductors exhibit similar Pair Distribution Functions (PDFs), characteristic of their underlying network-forming nature. On the other hand, amorphous metallic systems also display internally consistent PDF profiles, but different from those of the semiconducting materials. A comparative analysis of short-range and medium-range order reveals that while semiconductor structures feature a well-isolated first peak, with a zero-intensity region between the first and second peaks, metallic systems maintain a significant non-zero value between the first and second peaks. Furthermore, the second peak in metallic systems is bimodal, featuring a distinct elephant-like profile. Amorphous semi-metals display a still different profile, and the PDFs for Bi, for example, are similar to those for As and Sb. To deepen this structural comparison, we have incorporated amorphous Plane Angle Distributions (PADs), providing a more complete perspective on the local geometry. We introduce a renormalization approach that uses the positions of the first peaks in the PDFs to quantify these structural coincidences and discuss the implications.

## INTRODUCTION

The atomic-scale structure of a material is the cornerstone of modern materials science, governing its fundamental properties. While crystalline solids benefit from well-established structure-property relationships enabled by long-range order, non-crystalline materials, such as glasses and liquids, present a formidable challenge due to their inherently disordered arrangements. Despite their technological importance in applications ranging from semiconductors in computer science to metallic glasses in mechanical manufacturing, the absence of translational symmetry in these materials has long hindered a unified structural theory. However, advances in local-structure characterization techniques, particularly Pair Distribution Functions (PDFs) and Plane Angle Distributions (PADs), now provide a powerful lens to decode the short- and medium-range order that defines these systems.

Over the past two decades, our research group has systematically investigated the atomic structure of diverse non-crystalline materials through *ab initio* molecular dynamics (AIMD) simulations. The work began with covalent semiconductors, including amorphous carbon, silicon, and germanium[1], and was expanded to include metallic glasses (e.g., Al[2], Ag[3], Au[4], Cu[5–7], Pd[8–11], and Pt[12]), then continued with semi-metals (As[13], Bi[14–17], and Sb[18]). This extensive dataset has revealed intriguing structural patterns that transcend individual materials. Notably, PDFs and PADs, which quantify the probability of atomic pair distances and the distribution of plane angles, respectively, exhibit class-specific signatures in their first and second coordination shells and in the presence of certain angles preponderantly over others. For example, covalent amorphous solids display pronounced first-peaks, linked to bond-angles corresponding to tetrahedral bonding, while metallic glasses show asymmetric second peaks indicative of medium-range packing motifs. These features correlate strongly with macroscopic behaviors such as mechanical stability, glass-forming ability, and electronic properties.

Despite these insights, a systematic exploration of structural commonalities across distinct classes of non-crystalline solids remains elusive. Do universal principles underlie this apparent structural diversity? Can subtle similarities in PDF peak profiles suggest predictive models for novel materials? This study addresses these questions by conducting a comparative analysis of PDFs and PADs for amorphous covalent, metallic, and semi-metallic systems. By identifying shared structural motifs, we establish a framework to classify non-crystalline materials based on their local structure. Furthermore, we demonstrate that once these PDFs are renormalized, notable hidden similarities emerge that bridge the gap between seemingly disparate material classes.

# METHOD

The *undermelt-quench* method[19] is a computational protocol specifically devised to generate physically realistic amorphous structures by circumventing the limitations of traditional melt-and-quench strategies[20–22]. While standard quenching from a high-temperature liquid can sometimes fail to capture the specific metastable configurations observed experimentally, the *undermelt-quench* approach systematically destabilizes a crystalline unstable precursor to kinetically arrest the system in a disordered solid state. The procedure is implemented through three interrelated stages, each meticulously designed to disrupt long-range periodicity while preserving the short-range atomic arrangements characteristic of non-crystalline solids.

## *1. Selection of a Structurally Unstable Crystalline Precursor*

The initial phase involves selecting a crystalline polymorph that exhibits intrinsic structural instability, such as a high-energy phase or a significantly strained lattice configuration. This inherent instability lowers the energy barriers associated with amorphization, facilitating the transition to a disordered state during subsequent thermal treatments. For elemental systems, we avoid the most stable ground-state structures (e.g., diamond cubic for Si or Ge, or FCC for Pd). Instead, we utilize high-pressure phases or metastable configurations that possess the desired coordination and density but lack long-range stability under ambient conditions. A thermodynamically stable configuration would inherently resist disordering and favor recrystallization during the cooling phase; by starting from a "frustrated" precursor, we bypass these crystalline traps.

To ensure physical validity, the system density is constrained to align with experimental values of the corresponding amorphous phase. For instance, in the case of amorphous carbon, density is a critical parameter that dictates the $sp^2/sp^3$ bonding ratio. Similarly, for metallic systems like Au or Pd, maintaining the experimental amorphous density ensures that the short-range packing motifs are captured accurately. In the absence of direct experimental data for specific amorphous states, densities from metastable crystalline analogs serve as practical approximations, though these are always subjected to post-simulation validation against known structural benchmarks.

## *2. The ab initio Molecular Dynamics (AIMD) Simulation Process*

All *ab initio* calculations for this work were performed utilizing the CASTEP[23–25] and the DMol³ codes[26–29]. The electronic exchange-correlation interactions were treated within the Local Density Approximation (LDA) utilizing the Vosko-Wilk-Nusair (VWN) functional[30]. Core electrons were modeled using DFT Semi-core Pseudopotentials (DSPP)[29], while valence electron wavefunctions were expanded using a Double Numerical plus Polarization (DNP) basis set with a global orbital cutoff radius of 4.5 Å. To accurately handle orbital occupancies across semiconductor, semi-metallic, and metallic systems, Fermi smearing was applied, and Brillouin zone sampling was restricted to the Gamma-point. A rigorous self-consistent field (SCF) density convergence threshold of $1.0 \times 10^{-6}$ was enforced.

The *AIMD* protocol encompasses two dynamically distinct phases designed to transition the system from a frustrated crystalline state to a metastable amorphous configuration:

- **Heating (Disordering Phase):**
  The system is gradually heated from 300 K to a temperature precisely tuned to sit just below its equilibrium melting point, $T_m$ (e.g., 1650 K for Silicon, where $T_m$=1687 K). This "*undermelt*" temperature is strategically selected to provide sufficient thermal energy to break rigid crystalline atomic bonds and disrupt long-range crystalline symmetry while preventing the complete loss of local structural correlations. During the heating steps of the simulation (typically 1.0-2.0 ps), the system evolves under an NVT ensemble controlled by a Nosé-Hoover thermostat.[31,32] Increased atomic displacements and bond-breaking events systematically destabilize the lattice. By remaining below the formal melting threshold, we significantly mitigate the risk of liquefaction, a prevalent issue in traditional melt-and-quench simulations.
- **Quenching (Kinetic Trapping)**:
  Subsequently, the system undergoes rapid cooling from the sub-melting temperature to about 0 K over roughly 200 steps (~2.0–4.0 ps). This ultrafast quenching rate (~$10^{13}$–$10^{14}$ K/s) emulates kinetic arrest mechanisms found in experimental rapid-solidification techniques, such as splat cooling or melt-spinning. The rapid temperature reduction effectively "locks" the atoms into local minima of the potential energy landscape, stabilizing disordered configurations before they can revert to a crystalline state. In semiconductors, this stage yields the characteristic sharp, well-separated first peaks in the PDFs, with a coordination number (CN) just below 4. Unlike the melt-and-quench methods that preserve some of the liquid characteristics with a CN higher than 4. In pure metals, it preserves the dense, tetrahedral packing responsible for the bimodal second peak.

### *3. Geometry Optimization (GO): Energy Minimization*

Following the *AIMD* simulation process, the structure undergoes rigorous geometry optimization through the BFGS energy minimization method[33–36], utilizing the same electronic parameters (VWN functional, DSPP pseudopotentials, and DNP basis set) established during the molecular dynamics runs. This stage is essential to eliminate unphysical high-energy atomic configurations, such as overlapping electron densities or overstretched bonds that may arise from the finite nature of the *AIMD* simulation.

Spanning approximately 1000 optimization iterations, this procedure systematically relaxes atomic positions until strict convergence tolerances are satisfied. Specifically, optimizations proceed until the energy change falls below $1.0 \times 10^{-5}$ Ha, the maximum atomic displacement is under $5.0 \times 10^{-3}$ Å, and the maximum gradient (force) drops below $2.0 \times 10^{-3}$ Ha/Å (0.05 eV/Å). This refinement ensures that local atomic coordination environments, such as tetrahedral bond angles in semiconductors or the icosahedral packing in metals[6,7,37,38], adhere to fundamental quantum-mechanical criteria. By finding the nearest local minimum on the potential energy surface, we produce metastable amorphous systems suitable for the subsequent analysis through their PDFs and PADs.

# RESULTS AND DISCUSSION

In this section, we gather and present information for each of the three families of elemental amorphous materials mentioned above: Covalent semiconductors, glassy metals, and semi-metals. Most of the results reported are due to the work in our group during almost two decades and point to a more provocative behavior: Both PDFs and PADs for each of the three amorphous families are very similar.

## I. AMORPHOUS SEMICONDUCTING SYSTEMS

Amorphous semiconducting materials are some of the first substances studied computationally. The initial approach was to melt the corresponding crystal, and once liquefied, the substance was rapidly cooled through the solid-liquid transition; but having started from the liquid state, the material was somewhat metallic and therefore some of the liquid characteristics were kept in the transition. Earlier results mentioned a CN slightly larger than 4, which indicated the presence of some metallic remnants in the process.

### *i)* *Amorphous Carbon (a-C)*

The structural characteristics of amorphous carbon (*a*-C) are uniquely sensitive to system density, which dictates the competitive hybridization between *sp*$^2$ and *sp*$^3$ states. Analyses reveal that *a*-C can be categorized mainly into two distinct regimes: High-Density Amorphous Carbon (*hda*-C) and Low-Density Amorphous Carbon (*lda*-C).

- High-Density Amorphous Carbon (*hda*-C): At higher densities (approaching 3.5 g/cm³), the Pair Distribution Function (PDF) is dominated by a sharp first peak at approximately 1.56 Å (**Figure 1 (a)**). This distance corresponds to the primary C–C bond length in *sp*$^3$-hybridized environments, indicative of a diamond-like tetrahedral network. The structural rigidity of this phase is evidenced by a near-zero intensity at the first minimum (1.91 Å) after the first peak, signifying a lack of interstitial atoms and a well-defined first coordination shell.
- Low-Density Amorphous Carbon (*lda*-C): Conversely, at lower densities (near 2.0–2.5 g/cm³), the first peak shifts slightly toward 1.42 Å, reflecting the prevalence of *sp*$^2$ trigonal bonding similar to graphitic or fullerene-like motifs. In this regime, the second peak in the PDF becomes more complex, as the presence of ring motifs (5-, 6-, and 7-membered) introduces a wider variance in medium-range distances.

Beyond the first coordination shell, both forms of *a*-C exhibit a broad second peak centered near 2.55 Å. However, the local geometry differs significantly: in *lda*-C, the angular distribution features peaks at 107.55° and 117.13° (**Figure 1 (b)**), reflecting a hybrid environment with significant *sp*$^2$ character. In contrast, the *hda*-C profile shifts toward 111.57° and 121.01° (**Figure 1 (d)**), where higher density forces a local arrangement more tightly packed. This transition demonstrates how density acts as the primary tuning parameter for the geometric and electronic framework of the amorphous carbon network[1].

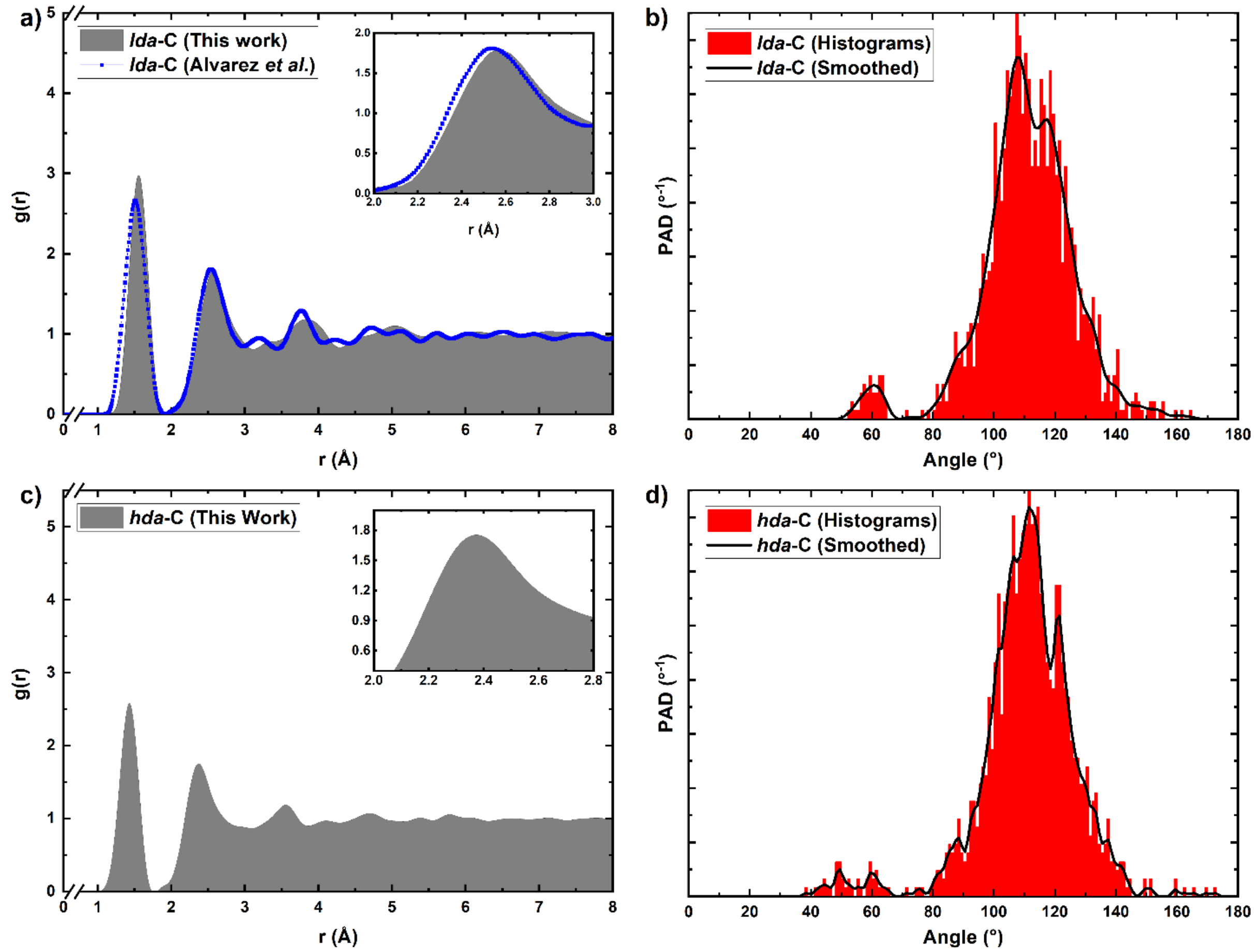

**Figure 1.** Structural evolution of amorphous carbon. (a) PDFs and (b) PAD for *lda*-C. (c) PDF and (d) PAD for *hda*-C. In (a), the 256-atom model (grey) is compared to the 64-atom model (blue dots) from Alvarez *et al.*[1] The insets in **Figures (a)** and **(c)** showcase the structure of the second peak.

### ii) Amorphous Silicon (a-Si)

In amorphous silicon (*a*-Si), the PDF provides information about its disordered tetrahedral network. The first coordination peak appears prominently at 2.36 Å (**Figure 2 (a)**), corresponding to the average Si–Si bond length. While the distance is slightly elongated compared to crystalline silicon (2.35 Å), the peak is broadened because of disorder that generates bond-angle distortions and coordination defects such as dangling bonds or floating bonds. Integration of this peak yields a CN close to 4.0, confirming the preservation of the local tetrahedral motif despite the loss of long-range periodicity.

A defining characteristic of the *a*-Si structure is the deep, near-zero minimum following the first shell, located at about 2.67 Å (**Figure 2 (a)**), as in *a*-C. This zero-valley indicates a rigorous spatial separation between the first and second neighbor shells, a hallmark of covalent network-formers that contrasts sharply with the filled interstitial spaces found in metallic glasses. Subsequently, a diminished and significantly broadened second coordination shell is evident between 3.0 Å and 4.5 Å, with a maximum at 3.78 Å. This peak is primarily governed by the distribution of Si–Si–Si bond angles; in a perfectly tetrahedral environment (109.5°) (**Figure 2 (b)**), the second neighbor distance is $r_2 \approx \sqrt{8/3}\, r_1$. The broadening observed here quantifies the angular structural disorder inherent to the amorphous state.

Beyond the second shell, the PDF oscillations dampen rapidly, signifying the absence of long-range order. However, a clear third peak can be observed between 5.0 Å and 6.0 Å. This third peak represents the medium-range order (MRO) and is sensitive to the topology of the ring motifs (typically 4-, 5-, and 6-membered rings)[39].

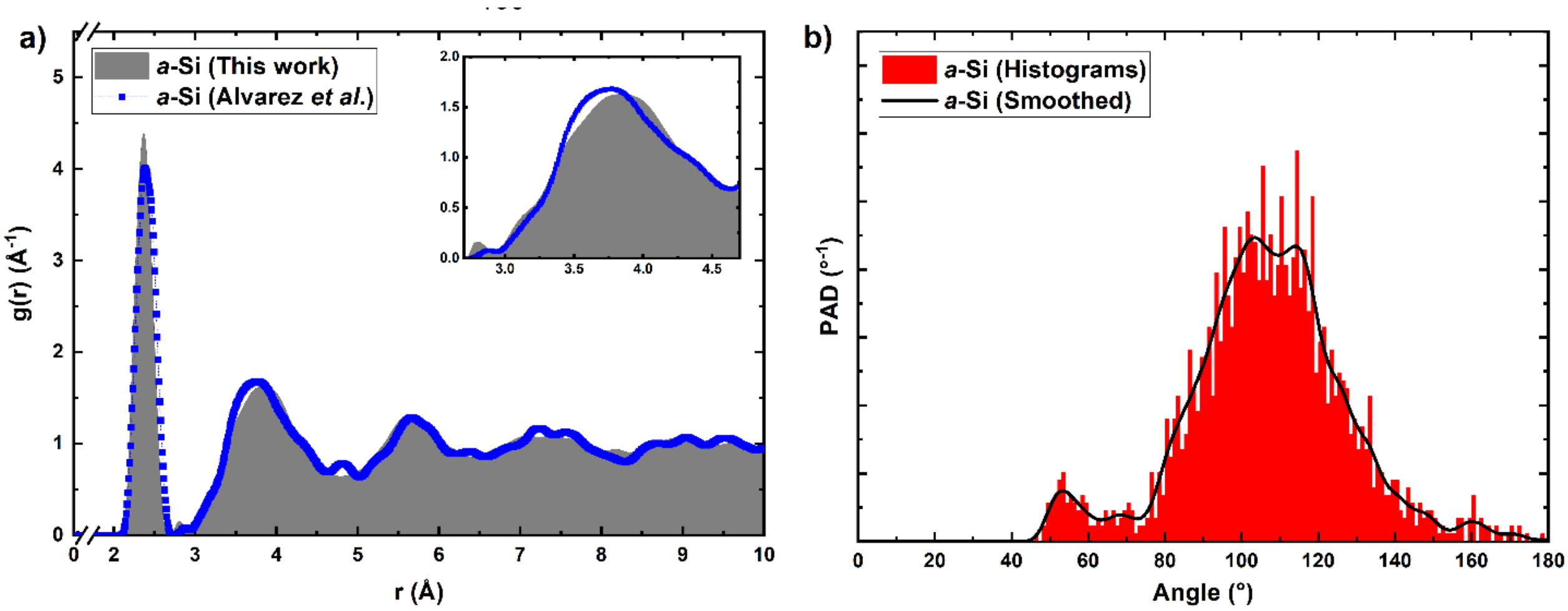

**Figure 2**. Structural characterization of amorphous silicon (*a*-Si). (a) PDFs and (b) PADs. In (a), the current 256-atom supercell (grey shade) is compared to the previously reported 64-atom model (blue dots) from Alvarez *et al.*[1]. The inset displays the differences between the two runs. (b) Two angles stand out, one at 60° and the other at about 110°

### iii) Amorphous Germanium (a-Ge)

The PDF analysis of amorphous germanium (*a*-Ge), shown in **Figure 3**, reinforces its classification as a prototypical semiconductor network. The structural profile exhibits a tall, narrow first coordination peak centered at 2.53 Å, representing the primary Ge–Ge covalent bond length, which in a crystal is 2.45 Å. This sharp feature indicates a high degree of short-range order, consistent with a somewhat deformed tetrahedral local environment.

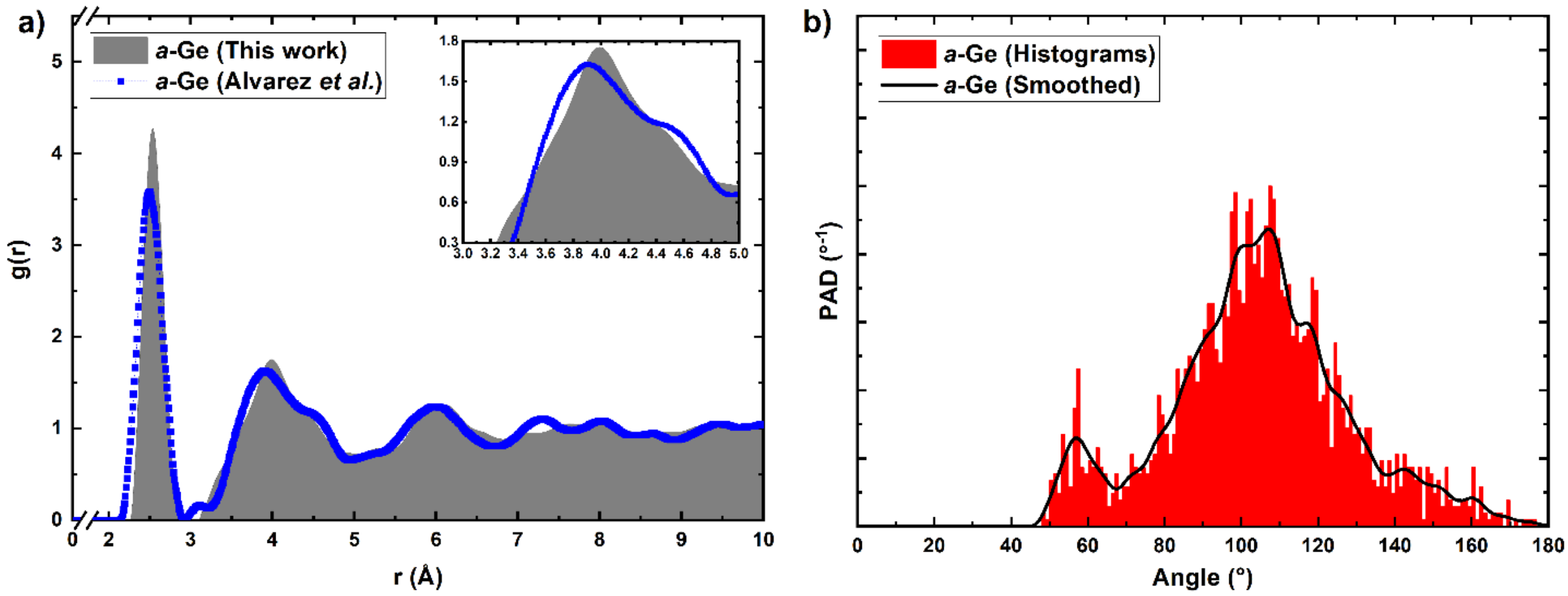

**Figure 3.** Structural characterization of amorphous germanium (*a*-Ge) showing the PDFs (a) and PADs (b) obtained. In (a), the current 256-atom supercell (grey shade) is compared to the previously reported 64-atom model (blue dots) from Alvarez *et al.*[1]. The inset displays the difference between the two works for the second peak. In (b), the angle distribution has a prominent peak at 110 ° and a smaller one at about 60 °.

A defining characteristic of this structural model is the presence of a clear zero-minimum at 2.93 Å (**Figure 3 (a)**). This absolute gap in the PDFs signifies a rigorous spatial separation between the first and second neighbor shells, confirming that, like *a*-Si, amorphous germanium maintains a continuous random network (CRN) devoid of interstitial atoms. The second coordination shell follows with a broad peak, reflecting the angular distribution of the tetrahedral bonding. The emergence of this well-defined "zero-valley" at 2.93 Å is a critical benchmark; it distinguishes the semiconductor class from the metallic and semi-metallic systems discussed later, where the first and second shells begin to overlap.

The PADs of **Figure 3 (b)** display two peaks, one at approximately 110°, prominent, indicating the presence of deformed tetrahedral angles; the other at 60°, indicating the presence of some equilateral triangles in the structure. The extension of the angular distribution reaches 180 °, starting at 50°. This spread speaks of an ample variety of angular motifs in the material.

## Comparative Analysis of Amorphous Semiconductors

The structural profiles of amorphous Carbon (*lda*-C and *hda*-C), Silicon (*a*-Si), and Germanium (*a*-Ge) reveal a remarkable consistency that defines the semiconductor class. Despite the significant differences in their atomic radii and their electronic structures, these amorphous materials share three fundamental structural "fingerprints":

- **Strict Shell Separation and Inter-shell Vacuum**: As illustrated in **Figure 4 (a)**, all three systems exhibit a distinct **zero-minimum** following the first coordination peak. This absolute gap signifies a rigid exclusion zone where no atoms reside. This "structural vacuum" serves as a definitive benchmark of the network-forming class.
- **Geometric Scaling and Tetrahedral Remnants**: The ratio between the coordination shell positions and the first peak ($r_n/r_1$) remains consistently close to the ideal values for an undeformed tetrahedral network, as detailed in **Table 1**. Specifically:
    - The second peak position aligns with $\sqrt{8/3} \approx 1.63$.
    - A third characteristic distance aligns with $\sqrt{19/3} \approx 2.51$.

  These invariant ratios (**Figure 4 (a)**) confirm that the Continuous Random Network (CRN) model is universally applicable across these elements, regardless of the specific atomic species.
- **Angular Symmetry**: The Plane-Angle Distributions (PADs) shown in **Figure 4 (b)** are characterized by a dominant peak centered near 109.5° ($sp^3$ coordination). However, a more detailed analysis reveals a more sophisticated angular landscape:
    - A shoulder peak at 116.57° indicates the presence of slight bond-angle distortions or residual $sp^2$-like planar correlations within the network.

- A small but clearly visible third peak at 60° signifies the existence of three-membered rings, for all the amorphous structure.

**Table 1.** Radial peak positions and geometric ratios for crystalline (*c*-) and amorphous (*a*-) semiconductors. Bond lengths ($r_1$) and subsequent neighbor shell positions ($r_2$, $r_3$, $r_4$) are given in Å. The dimensionless ratios $r_n/r_1$ illustrate the similarities of the renormalized curves, and the degree of adherence to the ideal tetrahedral scaling factors (**Figure 4 (a)**).

| | $r_1$ | $r_2$ | $r_3$ | $r_4$ | $r_2/r_1$ | $r_3/r_1$ | $r_4/r_1$ |
|---|---|---|---|---|---|---|---|
| *c*-C | 1.55 | 2.53 | 2.96 | 3.89 | 1.63 | 1.91 | 2.51 |
| *lda*-C | 1.58 | 2.58 | 3.18 | 3.88 | 1.63 | 2.01 | 2.46 |
| *hda*-C | 1.45 | 2.37 | 2.79 | 3.55 | 1.63 | 1.92 | 2.45 |
| *c*-Si | 2.35 | 3.85 | 4.51 | 5.92 | 1.63 | 1.91 | 2.51 |
| *a*-Si | 2.37 | 3.83 | 4.82 | 5.63 | 1.62 | 2.03 | 2.37 |
| *c*-Ge | 2.45 | 4.00 | 4.69 | 6.15 | 1.63 | 1.91 | 2.51 |
| *a*-Ge | 2.53 | 3.99 | 4.53 | 6.11 | 1.58 | 1.80 | 2.42 |

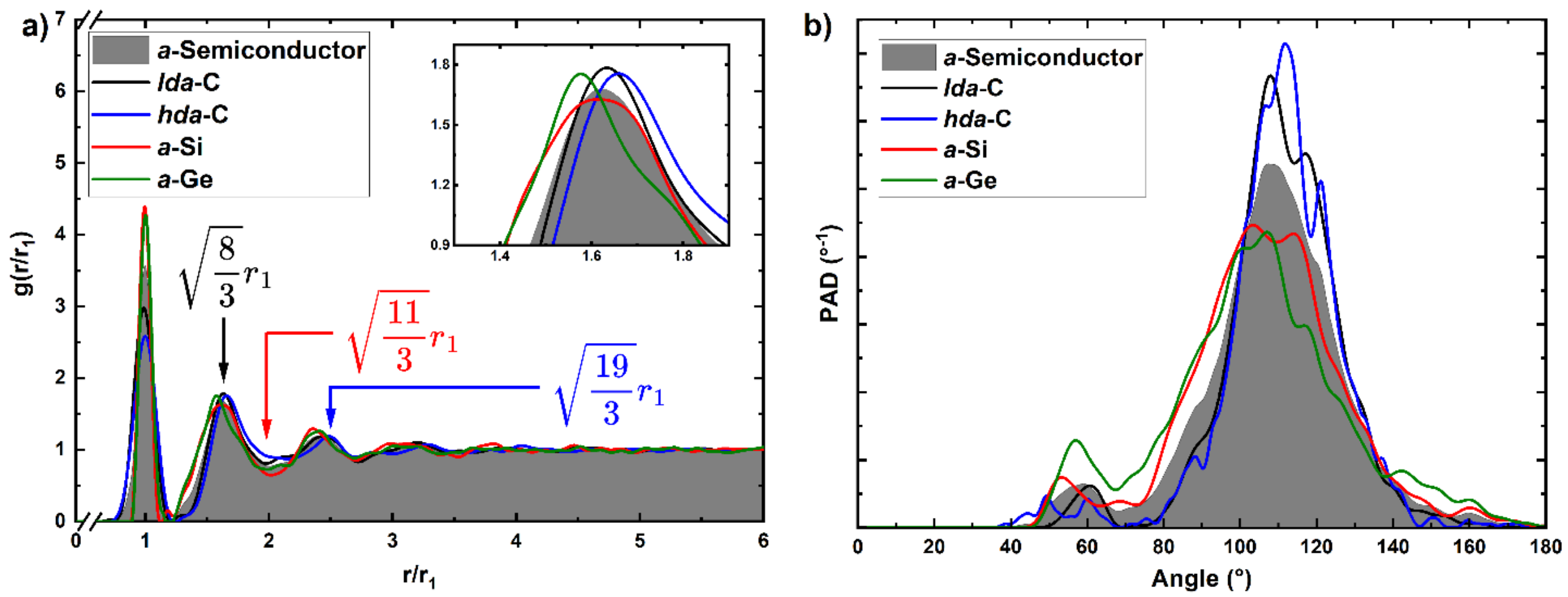

**Figure 4.** Comparative structural analysis of amorphous semiconductors (*lda*-C, *hda*-C, *a*-Si, and *a*-Ge). (a) Renormalized PDFs referred to the position of the first peaks and (b) PADs with two prominent peaks. This comparison illustrates the universal structural commonalities of the semiconductor class, characterized by distinct shell separation and tetrahedral angular distributions. The shaded gray area is the average of all others.

These commonalities suggest that amorphous semiconductors can be viewed as a single structural family. By applying a **renormalization factor** based on their primary bond lengths ($r_1$), their Pair Distribution Functions can be mapped onto a nearly identical master curve. This structural unity provides the necessary baseline to contrast the much more complex, densely packed configurations observed in amorphous metals and semi-metals.

## II. AMORPHOUS METALLIC SYSTEMS

The transition from amorphous semiconductors to amorphous metals represents a fundamental shift in the "rules" of atomic organization. In the previous section, we established that C, Si, and Ge act as structural architects; they utilize rigid, directional bonds to build open, covalent tetrahedral networks that enforce a strict spatial separation between the first two coordination shells. In these covalent systems, the "zero-minimum" in the PDF is a signature of a well-ordered emptiness between the first two shells.

In contrast, metallic systems, such as Al, Cu, Ag, Au, Pd, and Pt that we present below, abandon these directional covalent networks in favor of isotropic, short- and medium-range interactions that prioritize spatial efficiency above all else. Here, the structural narrative is no longer defined by what is absent, but by the crowded complexity of the medium-range order. As we move beyond the "zero-minimum" benchmark, we observe a "filling" of the inter-shell regions. This manifestation of metallic bonding creates a continuous landscape of structural correlations where the first and second coordination shells overlap. In these materials, the commonalities are found in the subtle, shared symmetries of the close-packing, a stark departure from the airy, deformed tetrahedral frameworks of their semiconducting counterparts.

### *i) Amorphous Aluminum (a-Al)*

Amorphous Aluminum serves as a primary example of the dense-packed metallic structural class. The PDF of *a*-Al (**Figure 5 (a)**) exhibits a narrow and pronounced first peak at 2.79 Å, which closely agrees with the metallic crystalline bond distance of 2.87 Å. In sharp contrast to the semiconductor group, the region immediately following this peak does not drop to zero; instead, it maintains a significant residual intensity, indicating a lack of a clear exclusion zone between coordination shells.

The second coordination shell, spanning the range of 4.5 Å to 5.7 Å, displays the characteristic bimodal "elephant peak" profile. This consistent feature, resembling an elephant, in the amorphous metallic systems, pays homage to the imaginative representation in Antoine de Saint-Exupéry's Le Petit Prince[40]. This feature is defined by two distinct maxima at 4.83 Å and 5.49 Å in this case. This splitting is a structural signature of amorphous metals, typically associated with the presence of competing local symmetries, of the close-packing, which are far more complex than the simple tetrahedral coordination found in covalent solids.

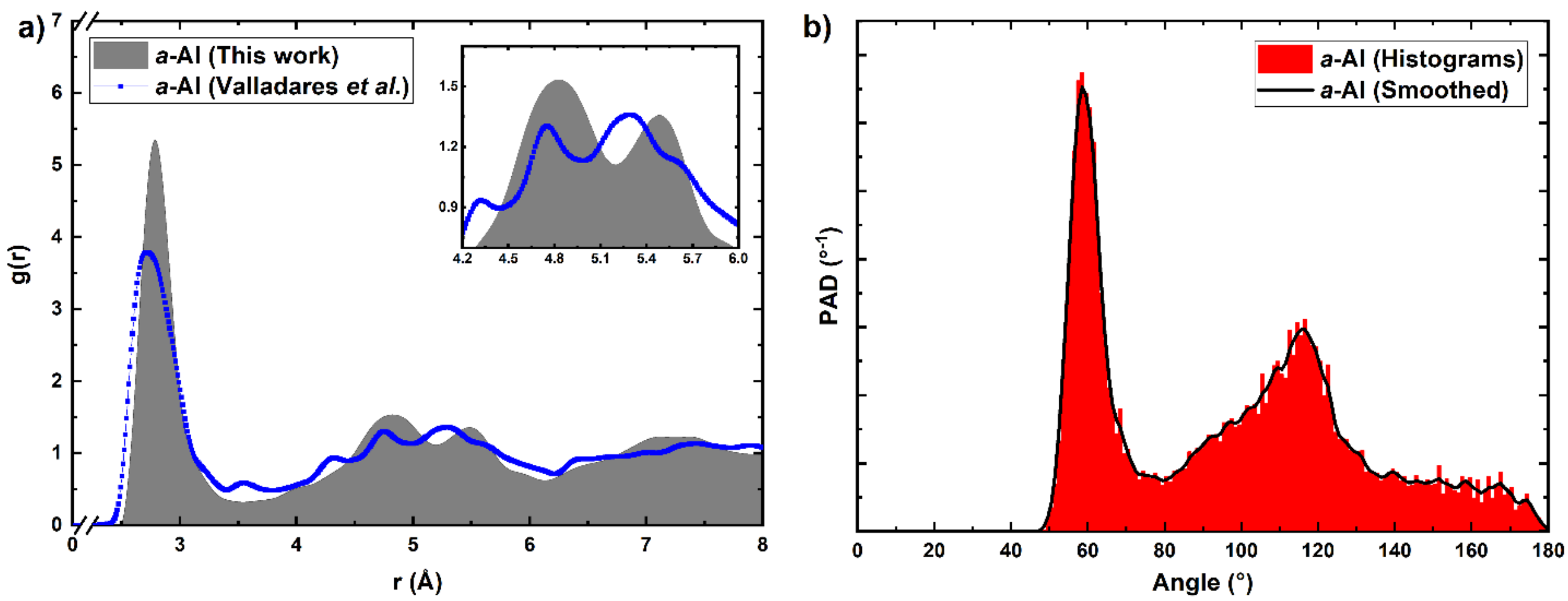

**Figure 5**. Structural characterization of amorphous aluminum (a-Al). (a) PDFs and (b) PADs. In (a), the current 216-atom supercell (grey shade) is compared to the 64-atom model (blue dots) from Valladares[2]. The inset shows the discrepancies between the two runs.

Further insight into the local geometry is provided by the PADs (**Figure 5 (b)**). The PAD for *a*-Al is characterized by a dominant, sharp peak at 58.5°, which corresponds to the nearly equilateral triangular faces of the icosahedral units that constitute the metallic glass. This is accompanied by a secondary, broader peak centered at 116°. The prominence of the 58.5°, as well as the substantial reduction in the peak at 116°, confirms the departure from the 109.5° tetrahedral angle of semiconductors, pointing instead toward a structure defined by the close-packing of atomic spheres.

### *ii) Amorphous Copper (a-Cu)*

Amorphous Copper further exemplifies the structural commonalities of the metallic class. As illustrated in **Figure 6 (a)**, the PDF of a-Cu is characterized by a tall and narrow first peak at 2.51 Å, which defines the primary metallic coordination distance in the amorphous, to be compared with the crystalline one of 2.55 Å. Similar to the other metals in this study, the region following this peak remains notably non-zero, reflecting the lack of a rigid exclusion zone and the presence of atoms in intermediate interstitial positions.

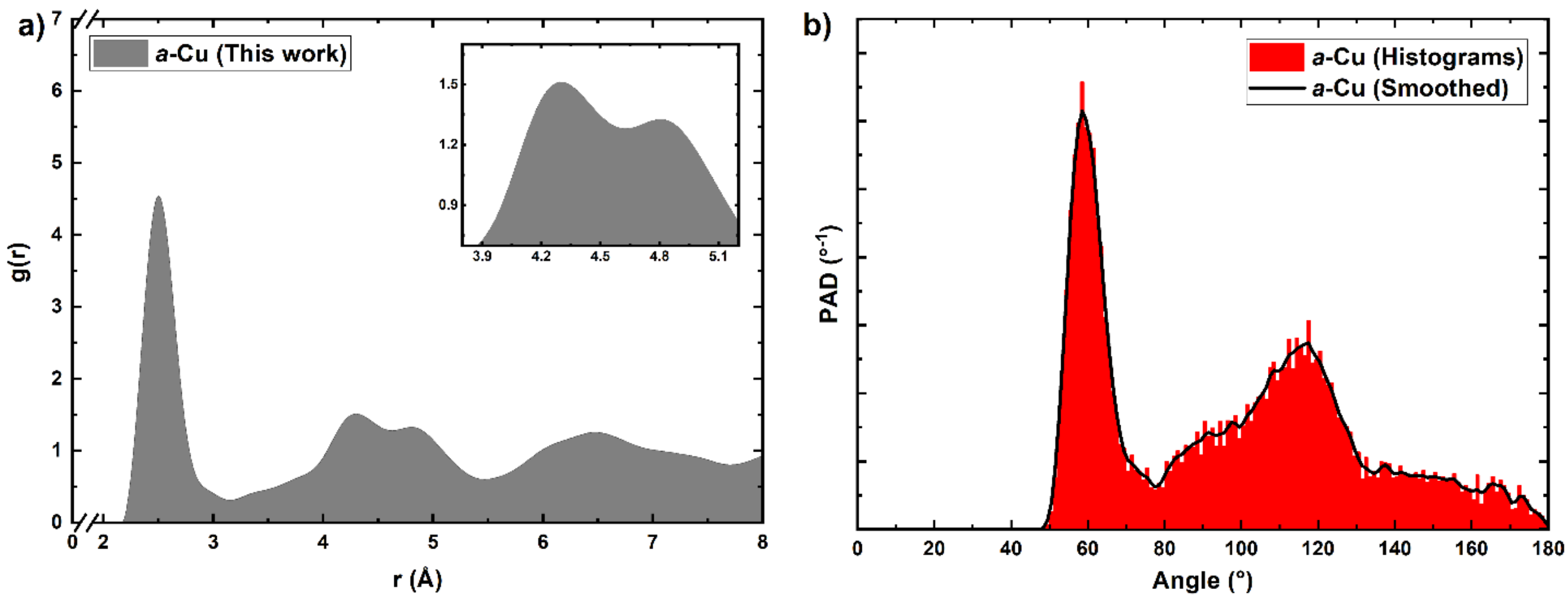

**Figure 6**. Structural characterization of amorphous copper (*a*-Cu). (a) PDF and (b) PADs. The elephant peak, characteristic of the amorphous metals studied, can be observed (Inset in **Figure (a)**)

A key indicator of the amorphous nature of this system is the departure from crystalline coordination. While crystalline copper (*c*-Cu) has a face-centered cubic (FCC) structure with a CN of 12, the amorphous phase settles into a slightly reduced average CN of 11.41. This coordination frustration is essential for preventing long-range crystallization and trapping the system in a disordered, metastable state.

The medium-range order in *a*-Cu is evidenced by the second coordination shell, located between 3.9 Å and 5.2 Å (**Figure 6 (a)**). This shell presents a well-defined bimodal elephant peak (see inset, **Figure 6 (a)**), with two distinct maxima at 4.29 Å and 4.82 Å. This doublet is a characteristic signature of the amorphous metallic state, arising from the complex, dense-packing motifs required to fill space efficiently without translational symmetry.

The PAD provides further insight into the local symmetry (**Figure 6 (b)**), featuring a narrow, dominant peak at 58.16°. This angle is consistent with the triangular faces of close-packed structures. Furthermore, the secondary feature is a prominent peak with a maximum at 116.9° and a short, narrow shoulder at 108.15°. The presence of this shoulder near the tetrahedral angle (109.5°), combined with the primary peak near the icosahedral angles (109.5° and 116.57°), suggests a complex structural landscape where distorted icosahedral units coexist with other polyhedral motifs. Moreover, there is a wide shoulder around 90° (at 91.04°) that does not seem to correspond to any of the icosahedral angles.

### *iii) Amorphous Silver (a-Ag)*

The structural profile of amorphous Silver reinforces the transition to dense-packed geometries, characterized by a narrow and tall first maximum at 2.81 Å (**Figure 7 (a)**) compared to the crystalline distance of 2.89 Å. As with aluminum, this first coordination shell is well-defined, yet the PDF remains notably non-zero in the subsequent valley, indicating a high degree of atomic packing efficiency that blurs the boundary between the first and second neighbor shells.

A critical distinction between the amorphous and crystalline phases of Silver lies in the local coordination. While crystalline Silver (*c*-Ag) maintains a perfect face-centered cubic (FCC) lattice with a CN of 12, our AIMD results show that *a*-Ag settles into an average CN of 11.34. This reduction reflects the presence of voids in the first coordination shell and the geometric frustration inherent in the disordered state, where the atoms cannot satisfy the ideal 12-fold packing of the crystal.

The second coordination shell, extending from 4.4 Å to 5.9 Å, presents a well-resolved bimodal elephant peak. The two maxima are located at 4.85 Å and 5.47 Å. The emergence of this doublet is a critical indicator of medium-range order in metallic glasses; specifically, the ratio of these sub-peaks is a fingerprint for the stability of the amorphous phase under rapid quenching.

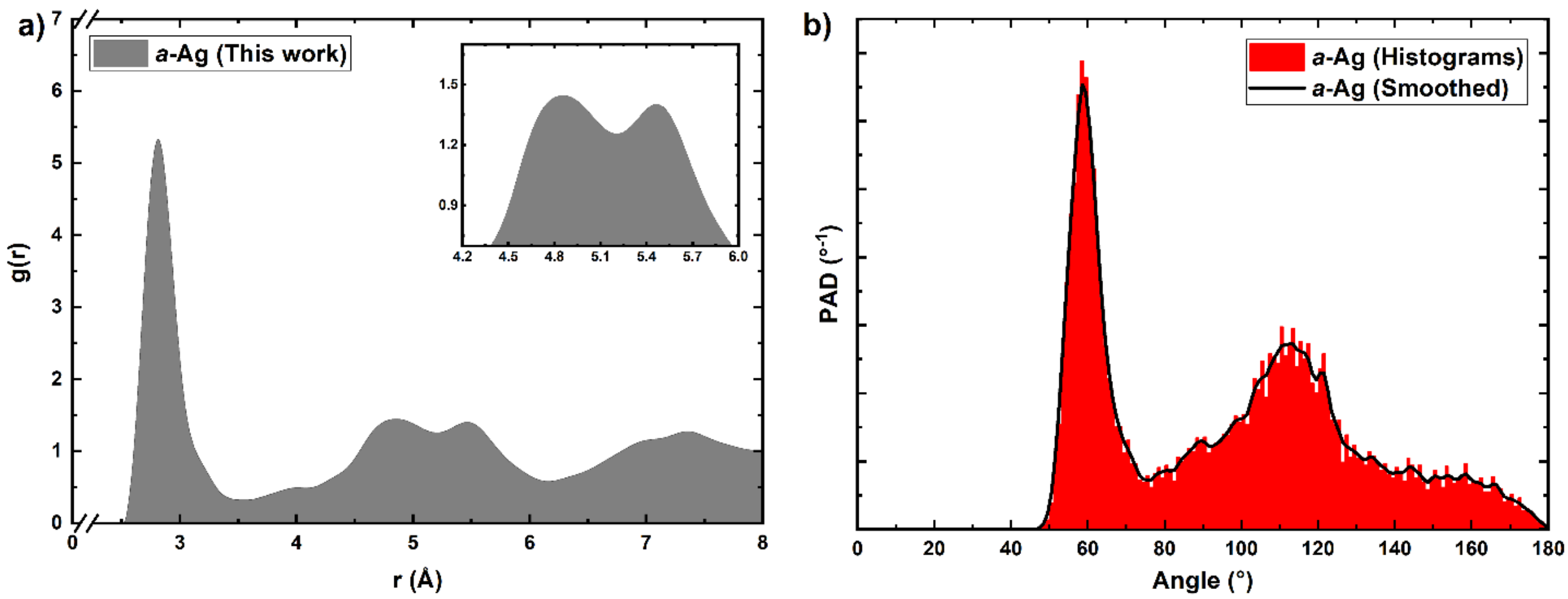

**Figure 7**. Structural characterization of amorphous silver (*a*-Ag). (a) PDF and (b) PADs. The bimodal structure of the second peak is displayed in the inset.

The PAD of *a*-Ag offers the most compelling evidence for its underlying symmetry. We observe a clear, tall, and narrow peak at 58.67° (**Figure 7 (b)**), which aligns with the triangular facets of most polyhedral structures. More interestingly, the secondary feature is a broad, complex peak with a primary maximum at 113.21° and a distinct shoulder at 120.72°.

These angular signatures are characteristic of icosahedral short-range order (ISRO). In an ideal icosahedron, the angles between atoms and the center of the cluster are near 63.43° and 116°, as will be demonstrated later. The shift and splitting observed here, particularly the shoulder at 120.72°, suggest distorted icosahedral units that are packed together to form the medium-range framework. This confirms that *a*-Ag, while sharing the same close-packing principles as aluminum, seems to exhibit a more pronounced tendency toward icosahedral clusters.

### *iv) Amorphous Gold (a-Au)*

Amorphous Gold serves as a definitive case-study in the structural behavior of heavy noble metals. As illustrated in **Figure 8**, the PDF of *a*-Au is defined by a tall, narrow first peak at 2.79 Å, which represents the primary metallic bond distance. When compared to the crystalline distance of 2.88 Å, this feature reveals the significant atomic repositioning required to sustain the disordered state. Consistent with the metallic family, the PDF exhibits a persistent non-zero valley following the first shell, a hallmark of high packing efficiency where the absence of a rigid exclusion zone allows for the overlap of local coordination environments. While crystalline gold (*c*-Au) adopts an FCC structure with a CN of 12, the amorphous phase settles into a lower average CN of 11.59. This value is the highest among the metals studied, reflecting a diminished degree of geometric frustration. Such a high coordination suggests that the amorphous gold network retains a strong structural proximity to the FCC phase, which correlates with its tendency toward recrystallization.

The medium-range order in *a*-Au is evidenced by a well-resolved bimodal elephant peak in the second coordination shell, extending from 4.5 Å to 5.8 Å (Inset of **Figure 8 (a)**). This shell features two distinct maxima at 4.91 Å and 5.51 Å. The emergence of this specific doublet is a universal signature of the amorphous metallic state, arising from the close-packing structures required to fill three-dimensional space in the absence of translational periodicity.

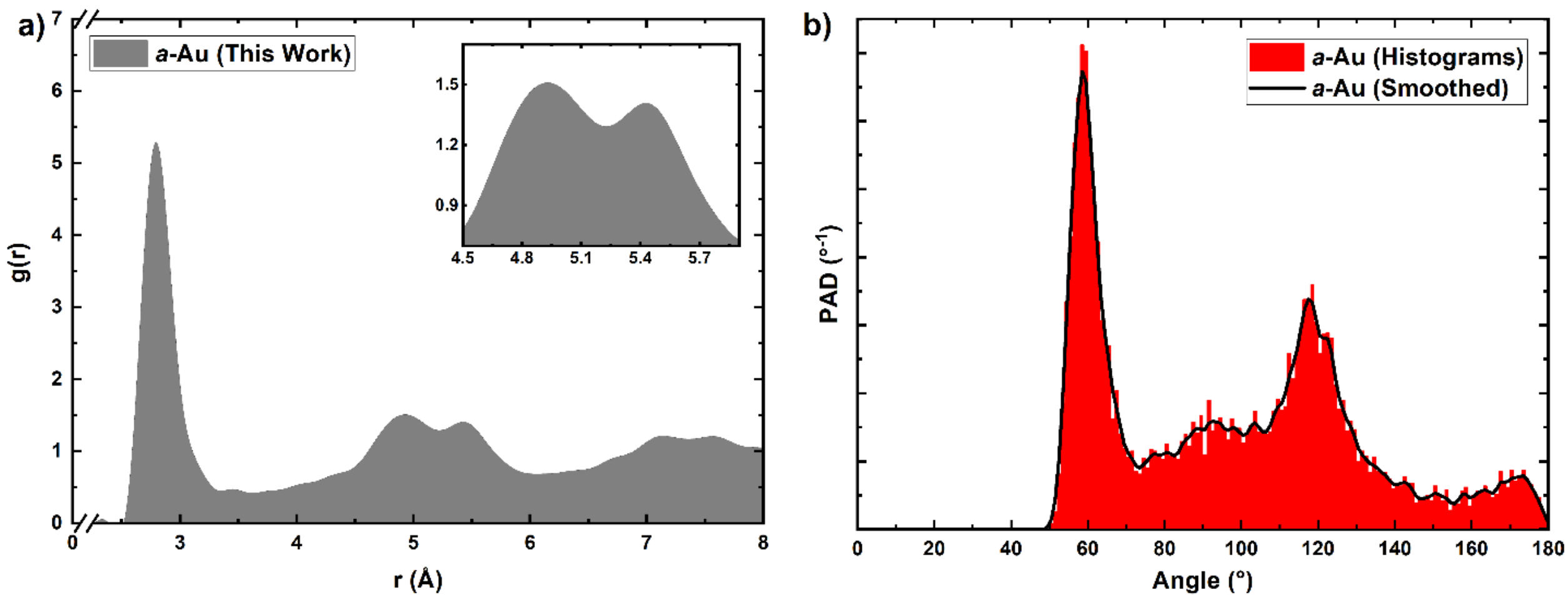

**Figure 8**. Structural characterization of amorphous gold (*a*-Au). (a) PDF and (b) PADs. In the inset, the structure of the second peak is displayed.

The local symmetry of the *a*-Au system is most clearly revealed through the PAD shown in **Figure 8 (b)**. While most amorphous metals in this study are dominated by the icosahedral signatures of close-packed clusters, Gold displays a distinct departure from this trend. The PAD of *a*-Au is characterized by a sharp primary peak at 59.21°, and a secondary peak located at 117.85°, but its most significant feature is a notable third peak at 90°. In the context of amorphous packing, a peak at this specific angle is a direct indicator of orthogonal or cubic-like local structures, mirroring the 90° plane angles found in the crystalline FCC lattice, a cubic memory.

Gold is difficult to amorphize precisely because it retains memory of the FCC structure. It crystallizes readily since it maintains some of the angle and topology features of the crystalline structure. The structure we used in this work was the one just before the crystallization process started, and is very similar to the ones reported for the noble-metal counterparts. Collectively, these PADs identify *a*-Au as a unique member of the metallic suite where the "amorphous" state is heavily influenced by crystalline structural memory.

### *v) Amorphous Palladium (a-Pd)*

Amorphous Palladium provides a look at the "crowded" structural landscape of metallic glasses. As shown in the PDF profiles of **Figure 9 (a)**, we analyzed three distinct configurations: *a*-Pd-I, *a*-Pd-II, and *a*-Pd-III, to ensure statistical robustness. The resulting average PDF (depicted in gray) (**Figure 9 (a)**) exhibits a sharp, tall first peak at 2.69 Å, closely resembling the crystalline metallic bond length of 2.75 Å. In sharp contrast to the semiconductor family, this coordination shell is immediately followed by a persistent non-zero valley. This lack of a clear exclusion zone indicates a high degree of atomic packing efficiency where the first and second shells overlap, a direct result of the structural rearrangements characteristic of the amorphization process.

A critical distinction in the local environment is the average CN of 11.10, the second lowest of the studied amorphous metals. While crystalline Palladium (*c*-Pd) maintains a perfect FCC lattice with a CN of 12, the amorphous state settles into a frustrated 11.10-fold arrangement. This reduction reflects the "free volume" created when atoms are displaced from equilibrium sites to occupy previously vacant interstitial spaces[41,42], thereby stabilizing the disordered configuration.

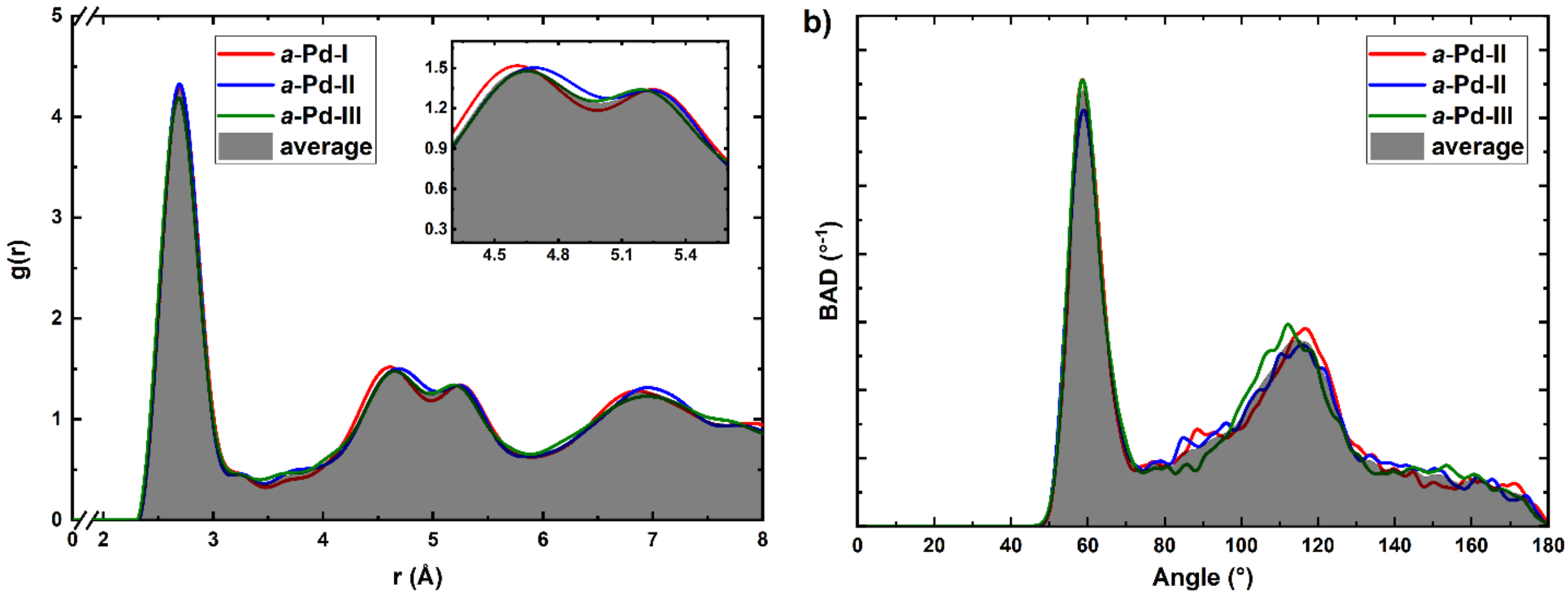

**Figure 9**. Structural characterization of amorphous palladium (*a*-Pd). (a) PDFs and (b) PADs. The profiles display results from three independent computational runs (*a*-Pd-I, **II**, and **III**), with the average represented by the gray shaded area. The inset shows the structure of the second peak for the three runs.

The medium-range order in *a*-Pd is characterized by the distinctive bimodal second peak, the elephant peak. In *a*-Pd, this doublet signifies the complex variations in local coordination required to maintain a stable disordered state.

The PADs (**Figure 9 (b)**) further confirm this metallic identity. The average PAD features a narrow and tall first peak at 58.8°, corresponding to the triangular facets of close-packed polyhedra. This is followed by a short, wide second peak with a maximum at 112.83° and a shoulder at 117.98°. These angular signatures demonstrate that *a*-Pd achieves spatial efficiency through a diverse and overlapping set of local symmetries, firmly establishing it within the metallic structural family.

### *vi) Amorphous Platinum (a-Pt)*

Amorphous Platinum further illustrates the "crowded" structural landscape characteristic of high-density metallic glasses. As shown in the PDF profile of **Figure 10 (a)**, *a*-Pt exhibits a sharp, tall first peak at 2.65 Å, which closely resembles the crystalline metallic bond length of 2.77 Å. In sharp contrast to the semiconductor family, this coordination shell is immediately followed by a persistent non-zero valley. This lack of a clear exclusion zone indicates a high degree of atomic packing efficiency where the first and second shells overlap, a direct result of the structural rearrangements and interstitial occupancy inherent to the amorphization process.

A critical indicator of the amorphous state in this system is the average CN of 10.97, the lowest of the studied metals. While crystalline Platinum (*c*-Pt) maintains a perfect FCC lattice with a CN of 12, the disordered phase settles into a frustrated arrangement. This reduction reflects the "free volume" created when atoms are displaced from equilibrium sites[42], effectively stabilizing the disordered configuration against crystallization.

The medium-range order in *a*-Pt is characterized by the distinctive bimodal elephant peak (**Figure 10 (a)**, inset). The presence of this doublet serves as a universal signature of the amorphous metallic class, arising from the close-packing motifs that fill space in the absence of translational symmetry.

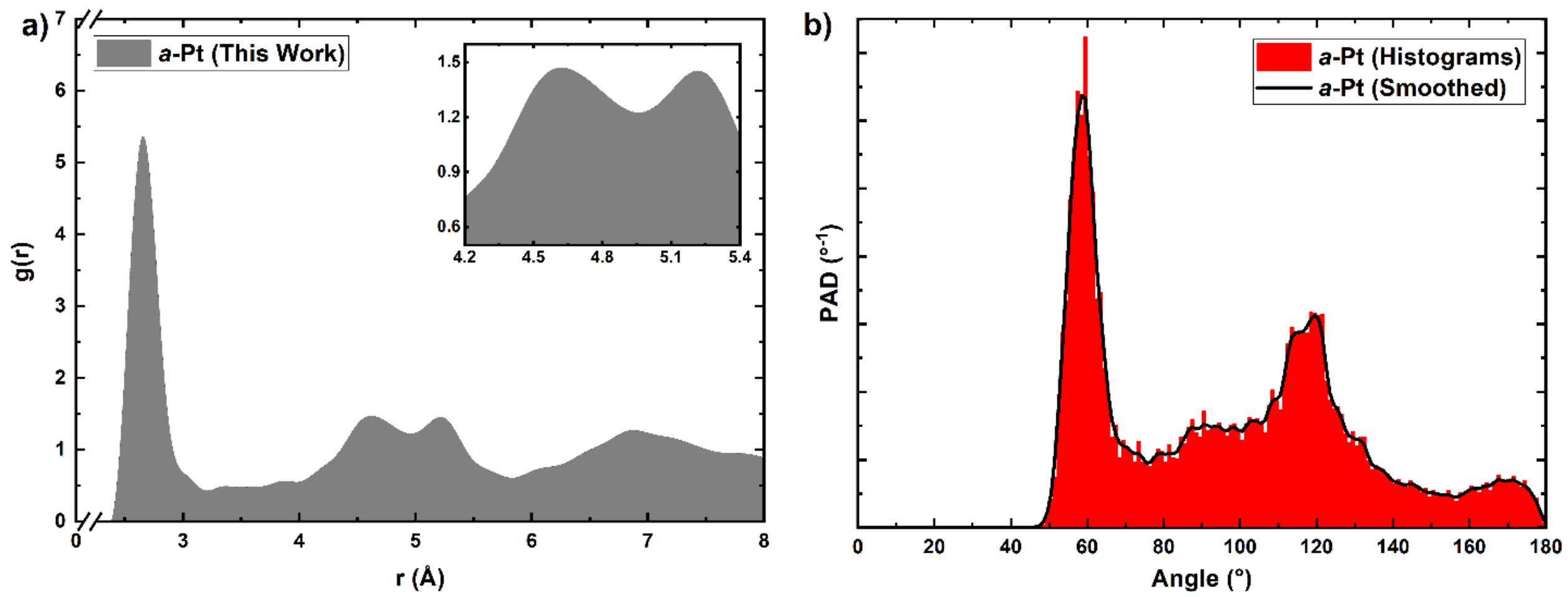

**Figure 10**. Structural characterization of amorphous platinum (*a*-Pt). (a) PDF and (b) PADs. The inset in **Figure (a)** displays the structure of the second peak

The PADs (**Figure 10 (b)**) further confirm this metallic identity. The PAD features a narrow and tall first peak at 58.16°, corresponding to the triangular facets of close-packed polyhedra. This is followed by a short, wide second peak with a maximum at 109.5°, which exactly matches the icosahedral angle in ISRO structures. These angular signatures demonstrate that *a*-Pt achieves spatial efficiency through a diverse and overlapping set of local symmetries, as *a*-Pd does, firmly establishing it within the metallic structural family alongside its noble metal counterparts.

## Comparative Analysis of Amorphous Metals

Just as the semiconductors share a structural identity, the metallic systems studied here: Al, Ag, Au, Cu, Pd, and Pt, exhibit a set of universal commonalities that define the metallic amorphous family. These shared features stand in stark contrast to the semiconductor "fingerprints" previously discussed:

- **Inter-shell Connectivity**: Unlike the "zero-minimum" seen in semiconductors, metals exhibit a persistent, non-zero valley between the first and second coordination shells **Figure 11 (a)**). As shown in **Figure 12 (a),** this region corresponds to the $\sqrt{2}\, r_1 \approx 1.41\, r_1$ position. This indicates a high atomic packing efficiency where intermediate interstitial sites are occupied, effectively blurring the boundary between neighbor shells.

- **The Elephant Peak (Bimodal MRO)**: A universal signature of medium-range order in the metallic glasses is the bimodal second peak in the PDF. Our analysis in **Table 2** confirms that this elephant peak is a composite of two distinct geometric motifs: a first maximum near $\sqrt{3}\, r_1 \approx 1.73\, r_1$ and a second near $2\, r_1$. This doublet reflects the competition between several Frank-Kasper structures[43,44], such as the icosahedral (Z12) and other Zs, required to fill space in the absence of a lattice.

- **Icosahedral Angular Dominance**: The PADs for these metals (**Figure 11 (b) and 12 (b)**) reveal a fundamental geometric shift, with a dominant maximum near 58.51° and a secondary feature at 116.51°. These angles closely track the ideal icosahedral benchmarks (60° and 109.5°), as shown in **Figure 12 (b)**, confirming that the local symmetry is driven by the triangular faces of close-packed atomic clusters.

- **Coordination Frustration and Scaling Invariance**: Despite varying atomic radii, the ratio of coordination shells ($r_n/r_1$) remains remarkably consistent across all studied amorphous metals (**Table 2**). In all cases, the average CN falls slightly below the ideal FCC value of 12 (typically between 10.9 and 11.6). This frustration is a mechanism that prevents recrystallization, though the higher CN in systems like a-Au (11.59) suggests a diminished barrier to ordering.

These commonalities suggest that amorphous metals can be unified under a single structural framework. By identifying these shared ratios, specifically the $1,\ \sqrt{2},\ \sqrt{3},\ 2,$ progression, we provide the empirical basis for the **renormalization approach**. This framework mathematically demonstrates the underlying symmetry connecting these seemingly disparate metallic elements into a single master curve.

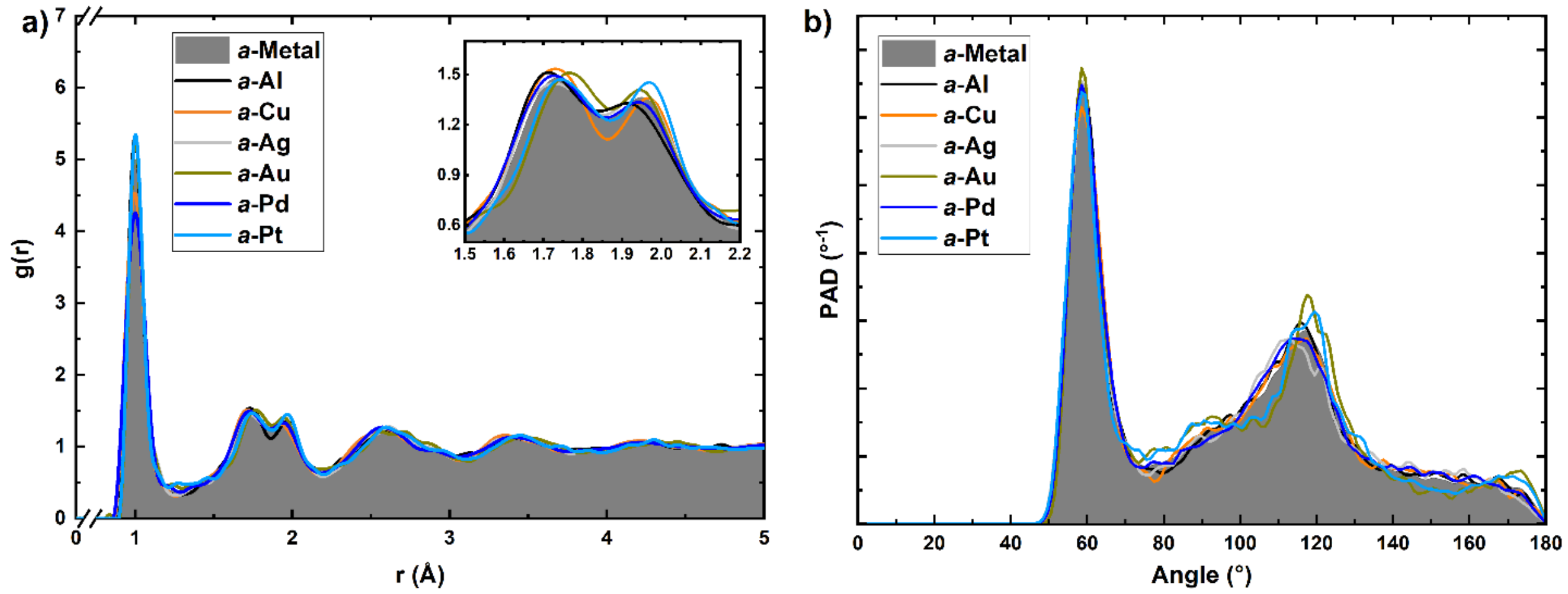

**Figure 11.** Comparative structural analysis of amorphous metals (*a*-Al, *a*-Cu, *a*-Ag, *a*-Au, *a*-Pd, and *a*-Pt). (a) Renormalized PDFs and (b) PADs. This comparison illustrates the universal structural commonalities of the metal family, characterized by an evident elephant peak. The shaded gray area is the average of all others.

**Table 2.** Radial peak positions and geometric ratios for crystalline (*c*-) and amorphous (*a*-) metals. Bond lengths ($r_1$) and subsequent neighbor shell positions ($r_2$, $r_3$, $r_4$) are given in Å. The dimensionless ratios $r_n/r_1$ illustrate the similarities of the renormalized curves, the Master curve.

| | $r_1$ | $r_2$ | $r_3$ | $r_4$ | $r_2/r_1$ | $r_3/r_1$ | $r_4/r_1$ |
|---|---|---|---|---|---|---|---|
| *c*-Al | 2.87 | 4.04 | 4.95 | 5.73 | 1.41 | 1.72 | 2.00 |
| *a*-Al | 2.79 | 3.93 | 4.83 | 5.49 | 1.41 | 1.73 | 1.97 |
| *c*-Cu | 2.55 | 3.62 | 4.43 | 5.12 | 1.42 | 1.74 | 2.01 |
| *a*-Cu | 2.51 | 3.69 | 4.29 | 4.82 | 1.47 | 1.71 | 1.92 |
| *c*-Ag | 2.89 | 4.09 | 5.01 | 5.78 | 1.42 | 1.73 | 2.00 |
| *a*-Ag | 2.81 | 4.01 | 4.85 | 5.47 | 1.43 | 1.73 | 2.04 |
| *c*-Au | 2.88 | 4.08 | 5.00 | 5.77 | 1.42 | 1.74 | 2.00 |
| *a*-Au | 2.79 | 4.11 | 4.91 | 5.51 | 1.47 | 1.76 | 1.97 |
| *c*-Pd | 2.75 | 3.89 | 4.77 | 5.50 | 1.41 | 1.73 | 2.00 |
| *a*-Pd | 2.69 | 3.79 | 4.65 | 5.23 | 1.41 | 1.73 | 1.95 |
| *c*-Pt | 2.77 | 3.93 | 4.80 | 5.55 | 1.42 | 1.73 | 2.00 |
| *a*-Pt | 2.65 | 3.85 | 4.53 | 5.21 | 1.45 | 1.74 | 1.97 |

The first coordination shell is centered at $r_1$ (blue), **Figure 12 (a)**. The "inter-shell" valley corresponds to the $\sqrt{2}r_1$ (green position), identifying atoms in intermediate interstitial sites. The characteristic bimodal peak is composed of two distinct geometric motifs: the first maximum near $\sqrt{3}r_1$ (purple) and the second at $2r_1$ (red). (b) For the average PAD, the angular distribution highlights a dominant peak at 58.51° and a secondary feature at 116.51°. The vertical red markers indicate the ideal angles of a perfect icosahedron (58.29°, 60°, 63.43°, 109.5°, and 116.5°), demonstrating that while the system is dominated by icosahedral close-packing, it retains significant structural distortion characteristic of the amorphous state.

As we descend the *p*-block of the periodic table, the elemental materials transition from the rigid covalent networks of semiconductors toward the metallic behavior of heavy elements. This electronic evolution is intrinsically linked to a fundamental structural transformation: the highly directional, covalent bonds that enforce tetrahedral symmetry begin to weaken, giving way to the less-directional, more-isotropic interactions characteristic of metals.

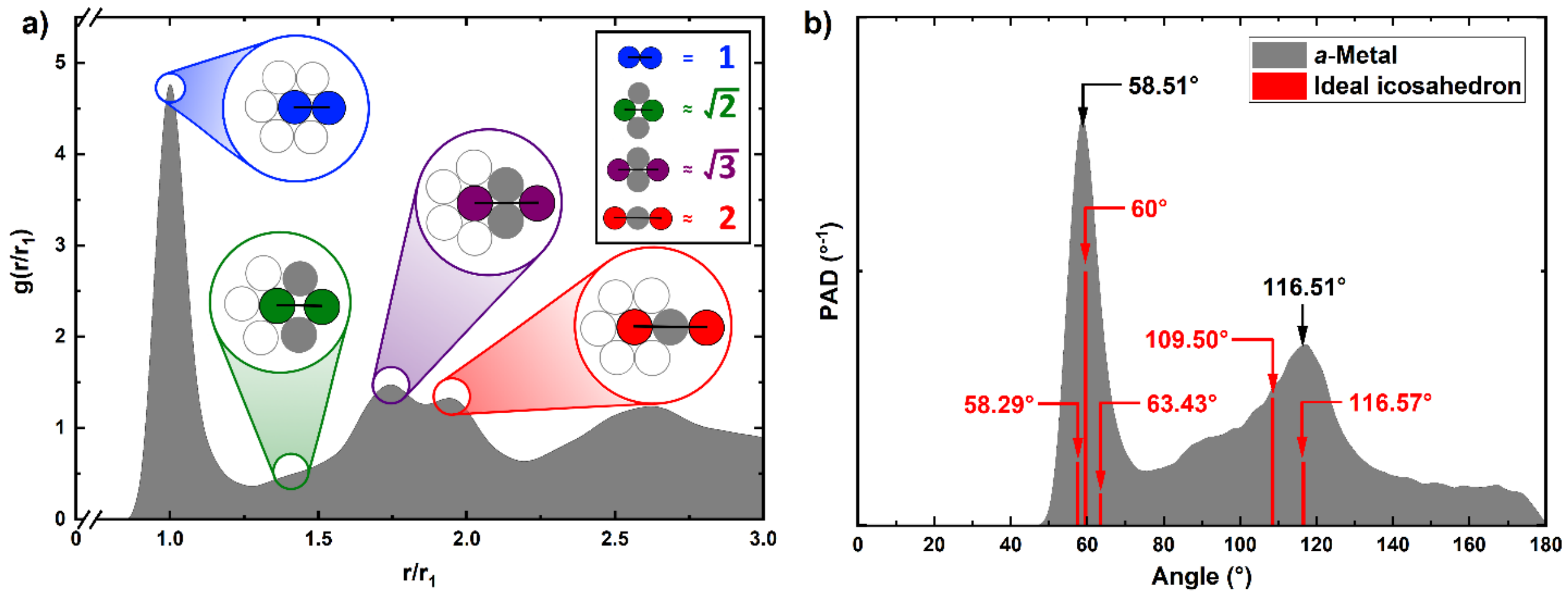

**Figure 12.** Structural benchmarks and local atomic environments in amorphous metals. (a) Normalized PDF: The profile illustrates in two dimensions the four primary Short- and Medium Range-Order regions identified across the metallic suite. (b) Comparison of average PADs (**Figure 11 (b)**) and the ideal angles of a perfect icosahedron.

## III. AMORPHOUS SEMI-METALLIC SYSTEMS

In the amorphous state, Arsenic (As), Antimony (Sb), and Bismuth (Bi) occupy a unique "structural bridge" in this progression. Their PDFs reflect a hybrid topology that is neither purely network-like nor entirely dense-packed. Their PADs, qualitatively similar, show an evolution from *a*-As to *a*-Bi that seems to indicate an imminent transition from a semiconductor behavior to a metallic behavior. One could speculate that if there were more semi-metals, this transition would become smooth from an amorphous semiconductor to an amorphous metal.

Another feature that emerges is that in passing from semiconductor-like behavior to metallic-like behavior, the zero-minimum valley of the semiconductors located between the first and second coordination shells starts disappearing. Atomic arrangements manifest that signal more complicated inter-shell structures, increasing the presence of atoms in this region.

### *i) Amorphous Arsenic (a-As)*

Arsenic represents the first stage of the semi-metallic transition, where the rigid tetrahedral exclusion zone of the semiconductors begins to destabilize. As shown in the PDF (**Figure 13 (a)**), the first coordination maximum is located at 2.49 Å, in agreement with the primary covalent bond length of 2.48 Å. However, unlike the "zero-minimum" observed in amorphous semiconductors, *a*-As exhibits a distinct inter-shell peak at 2.90 Å. This feature represents the local "filling" of the structural vacuum, indicating that the atoms are no longer strictly confined to a four-fold coordinated network and are beginning to occupy closer interstitial positions.

The medium-range order further reflects this transitional state. The second coordination shell at 3.74 Å lacks the singular sharpness of the CRN model, displaying instead a broader profile with an emerging shoulder located at about 3.5 Å. This fragmentation suggests a high degree of local coordination diversity, acting as a nascent precursor to the metallic elephant peak doublet (see inset of **Figure 13 (a)**).

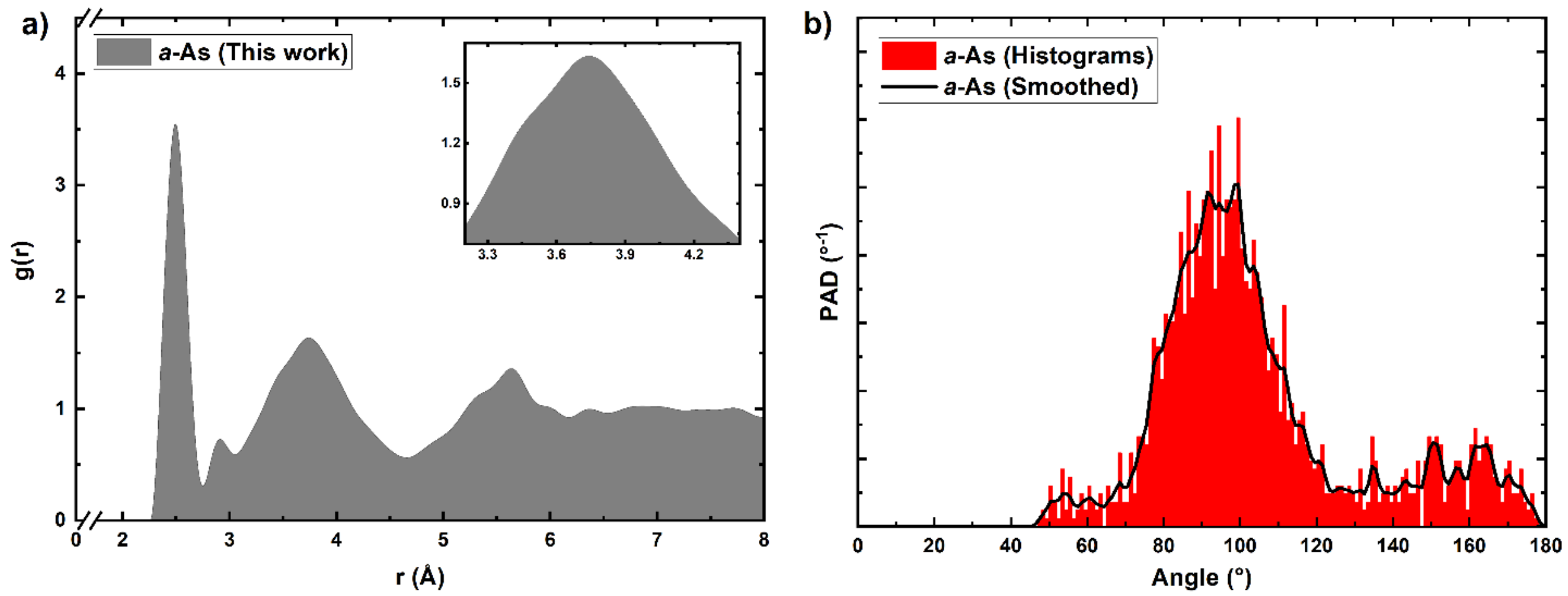

**Figure 13**. Structural characterization of amorphous arsenic (*a*-As). (a) PDF and (b) PADs. The inset in **Figure (a)** displays in more detail the structure of the second coordination shell.

The PADs (**Figure 13 (b)**) provide the most striking evidence of the semi-metallic shift. The distribution is dominated by a tall peak, with incipient bimodality; this bimodality has clear maxima at 91.3° and 99.0°. This strong preference for 90°, orthogonal symmetry, rather than the 109.5°, tetrahedral angle, highlights a move toward the *p*-orbital dominated bonding characteristic of the heavier group-15 elements. Furthermore, the appearance of a small peak between 45° and 60° signals the early formation of close-packed motifs. While much smaller than the primary peak, this feature represents the first geometric evidence of the metallic-like triangular facets that will become predominant as we descend further to Antimony and Bismuth. Angles larger than 120º also appear and indicate a diversification of the structures and of the angles, better associated with amorphous metallic systems. This feature appears both for As and Bi.

### *ii) Amorphous Antimony (a-Sb)*

Antimony represents a critical midpoint in the semi-metallic transition, where the distinction between coordination shells begins to blur. As shown in the PDF (**Figure 14 (a)**), the first coordination maximum is located at 2.97 Å. Crucially, the structural vacuum observed in semiconductors has completely collapsed, replaced by an incipient shoulder at 3.34 Å. This non-zero valley indicates that a significant population of atoms now resides in intermediate interstitial positions, as displayed in **Figure 12 (a)**, a direct consequence of the weakening of directional bonding constraints.

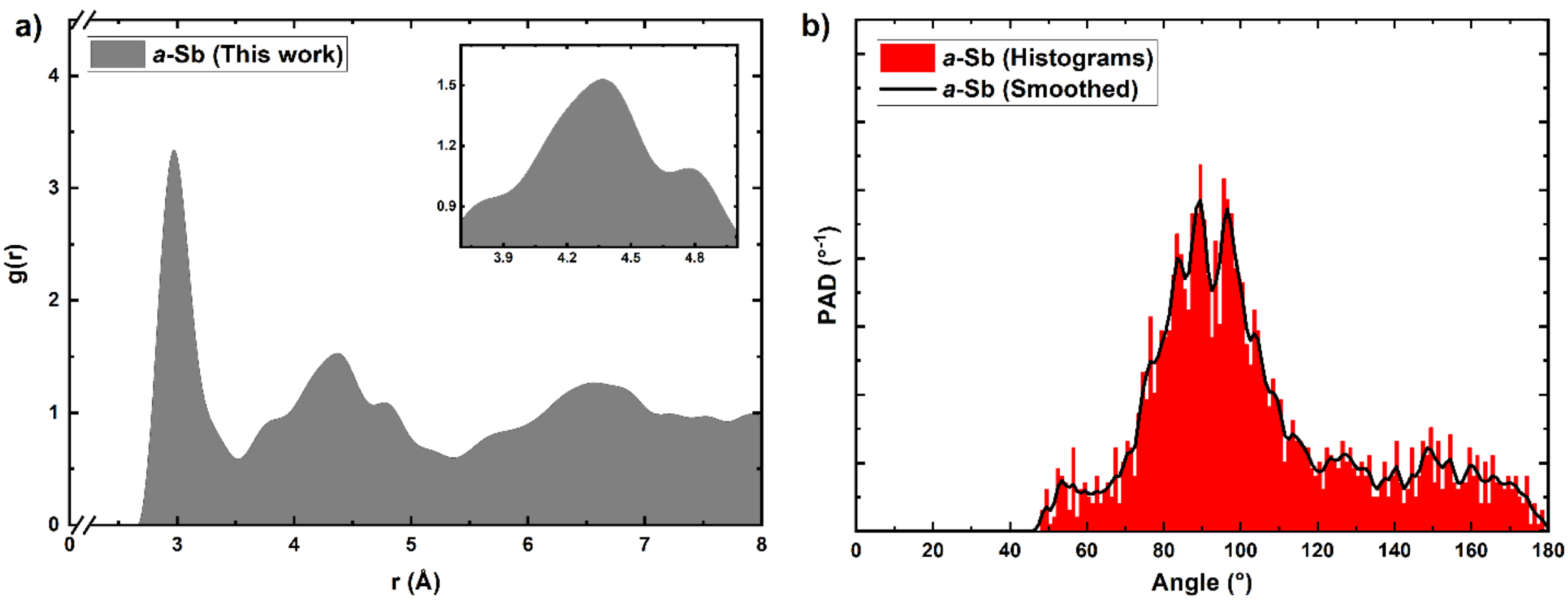

**Figure 14**. Structural characterization of amorphous antimony (*a*-Sb). (a) PDF and (b) PADs. The inset in **Figure (a)** displays, in more detail, the atomic structure

The medium-range order in *a*-Sb is notably more complex than in its lighter counterpart, Arsenic. The second coordination shell is no longer a single broad feature but is composed of three visible peaks at 3.79 Å, 4.37 Å, and 4.79 Å (see inset of **Figure 14 (a)**). This fragmented triplet is a sophisticated precursor to the metallic elephant peak, representing a highly frustrated state where multiple local packing features coexist but have not yet unified into the close-packed doublet seen in pure metals.

The PADs (**Figure 14 (b)**) reinforce this transitional identity. The distribution is dominated by a wide, tall primary feature with a triple-maxima signature at 83.32°, 89.20°, and 96.60°. This indicates a highly distorted orthogonal environment where possible *p*-orbital bonding still prevails, but is under increasing competition with metallic forces. Furthermore, the incipient peak at 53° is now more visible, signaling that the triangular facets of close-packed clusters are becoming a permanent fixture of the local topology. Also more visible are angles larger than 120º, which indicates a diversification of these features better associated with amorphous metallic systems.

### *iii) Amorphous Bismuth (a-Bi)*

As the heaviest stable semimetal in this family, amorphous Bismuth (*a*-Bi) represents the final structural stage before the complete transition to metallic behavior. The PDFs of *a*-Bi (**Figure 15 (a)**) are characterized by a primary first coordination maximum at 3.20 Å. Notably, this peak is broader than those observed in the pure metallic family, signaling a high degree of local bond-length fluctuations. The discrepancy between our present work and the one we performed several years ago, depicted in **Figure 15 (a)** (Inset)[17] we attribute it to the older version of the code we used then.

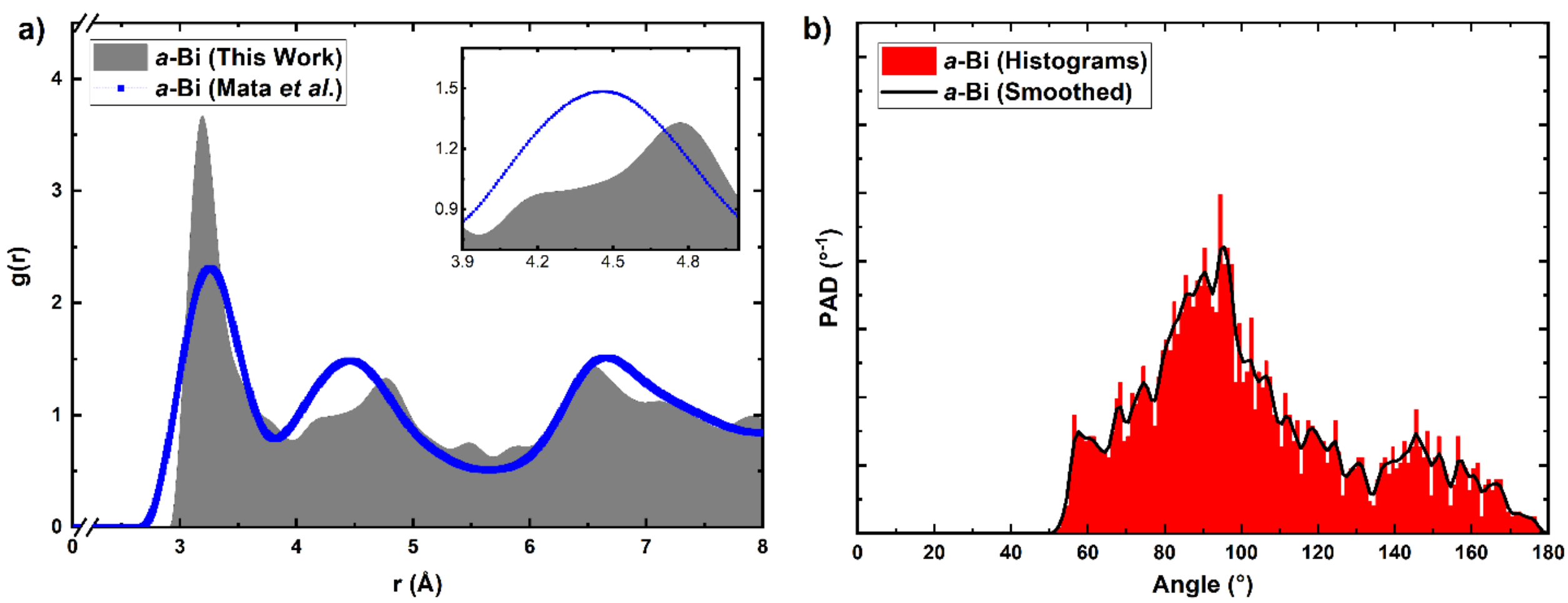

**Figure 15**. Structural characterization of amorphous bismuth (*a*-Bi). (a) PDFs and (b) PADs. In (a), the current 216-atom supercell (grey shade) is compared to the previously reported Mata *et al.*[16], and in the inset, the incipient bimodal peak appears.

Consistent with the metallic and semi-metallic transitions, the PDF of *a*-Bi exhibits a persistent non-zero minimum between the first and second coordination shells. This continuous atomic distribution indicates a substantial overlap of neighbor shells and the absence of a structural exclusion zone. Interestingly, the second coordination shell, centered at 4.77 Å, remains distinctly unimodal, but with a prominent shoulder near 4.20 Å, signaling the advent of the bimodal peak, the elephant peak. This broader, singular profile suggests that while the system is densely packed, it has not yet achieved the specific close-packed stability required to resolve the metallic MRO signatures.

The PADs provide the definitive evidence of Bismuth's transitional nature (**Figure 15 (b)**). Unlike the 109.5° tetrahedral peak of semiconductors or the dominant 60° peak of metals, the PADs for a-Bi are centered around a primary, wide peak at 90°. This indicates a dominant preference for orthogonal local geometries. This orthogonal dominance is accompanied by a secondary, lower-intensity peak at 55.27°. This feature represents a nascent, but significant, tendency toward the triangular facets of close-packed systems, marking the final geometric precursor to the close-packing of the metallic state. Angles larger than 120º also appear and indicate a diversification of the structures and of the angles better associated with amorphous metallic systems.

## Comparative Analysis of Amorphous Semi-metals

Interestingly, despite its similarities in the characteristics that identify them as another family of materials, the class of the semi-metallic elements clearly has attributes that indicate its transitional features in going from amorphous semiconductors to amorphous metals. The presence of a Master PDF for semi-metals, even though it exists, is more difficult to see. The gradual filling of the inter-shell vacuum is indicative of the increasing presence of atoms in this region, leading to characteristics more akin to the structure of amorphous metals. Summing up:

- **The Collapse of the Zero-Minimum of Amorphous Semiconductors**: Semi-metals exhibit a persistent, non-zero intensity between the first and second coordination shells, with an incipient shoulder to the right of the first peak, (around $1.2\, r_1$) and also a more prominent shoulder on the left of the second peak (around $1.3\, r_1$). The rigid exclusion zone seen in semiconductors begins to fill: A partial shell overlap.

- **Geometric Frustration of the Second Shell**: The medium-range order shifts from the single, sharp peak of the Continuous Random Network (CRN) toward a nascent, fragmented multimodality, the precursor to the metallic elephant peak. The presence of multiple shoulders (specifically in the 3.5–4.2 Å range) for Sb suggests a high degree of local coordination diversity and the early formation of the bimodal peak.

- **Angular Competition**: The Plane-Angle Distributions (PADs) reveal a competition between the 109.5° tetrahedral angle and the 60° close-packed angle, often mediated by 90° orthogonal motifs. The prominence of peaks near 90° (dominant in Sb and Bi) confirms that these systems are structurally frustrated from achieving the full close-packed structures of metallic glasses. However, the presence of a small peak at about 60º, and the widespread presence of angles beyond 120º presages the advent of structures closely related to amorphous metals.

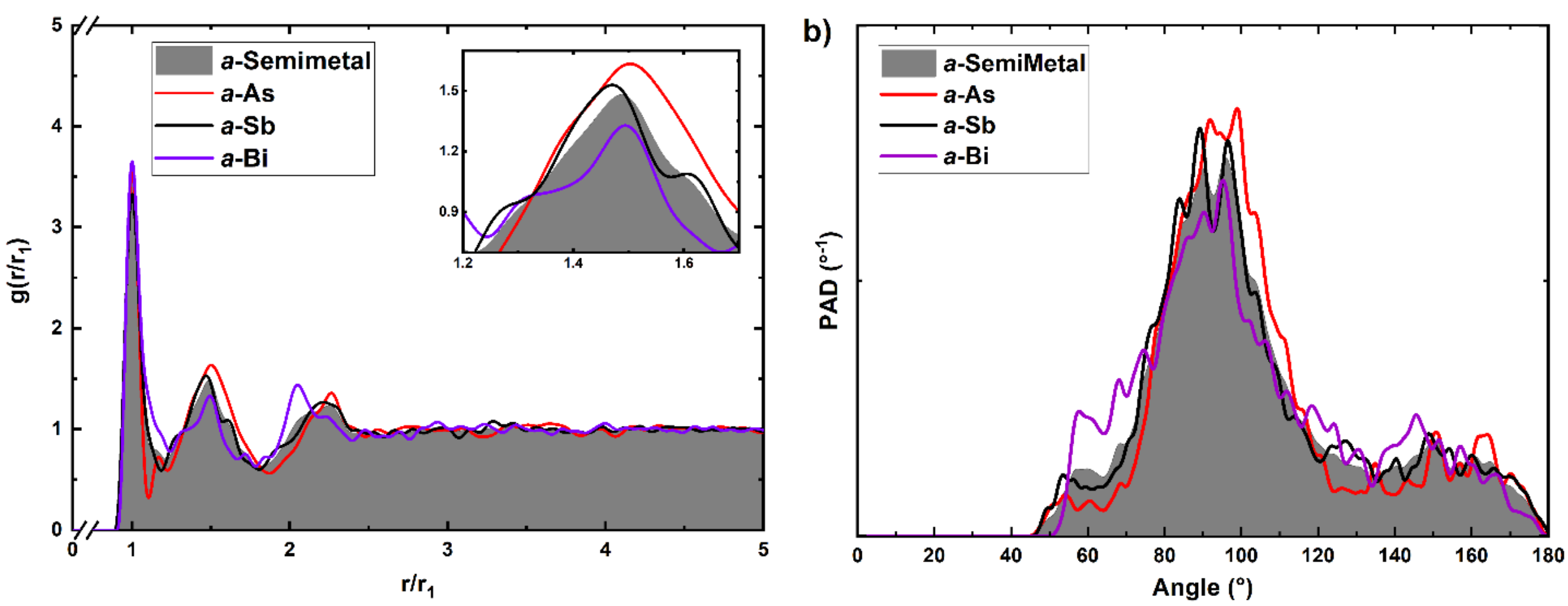

**Figure 16**. Structural unification of the semi-metallic family. (a) Renormalized PDFs and (b) PADs. The profiles illustrate the structural commonalities across the studied semi-metals (As, Sb, and Bi). The shaded gray area is the average of all others, and the inset in **(a)** registers the evolution of the bimodal peak towards a better-defined elephant peak.

The structural profiles of amorphous semi-metals, such as *a*-As, *a*-Sb, and *a*-Bi, define a transitional class that bridges the gap between rigid covalent networks and dense metallic packing, as can be seen in **Table 3**. While these systems exhibit a metallic-like lack of shell separation, they retain a clear geometric "memory" of their semiconducting precursors. Their universal commonalities include:

The individual runs for each element are displayed in **Figure 16**, with the average represented by the gray shaded area. The PDFs highlight the "filled" inter-shell valley, while the PAD showcases the characteristic broad orthogonal feature centered near 90°, marking the bridge between the tetrahedral and the close-packed geometries.

**Table 3.** Radial peak positions and geometric ratios for crystalline (*c*-) and amorphous (*a*-) semi-metals. Bond lengths ($r_1$) and subsequent neighbor shell positions ($r_2$, $r_3$, $r_4$) are given in Å. The dimensionless ratios $r_n/r_1$ illustrate the similarities of the renormalized curves, the Master curve.

| | $r_1$ | $r_2$ | $r_3$ | $r_4$ | $r_2/r_1$ | $r_3/r_1$ | $r_4/r_1$ |
|---|---|---|---|---|---|---|---|
| *c*-As | 2.42 | 3.94 | 4.71 | 5.58 | 1.63 | 1.95 | 2.31 |
| *a*-As | 2.49 | 3.74 | 5.36 | 5.64 | 1.52 | 2.15 | 2.27 |
| *c*-Sb | 2.84 | - | 4.64 | 5.41 | - | 1.63 | 1.91 |
| *a*-Sb | 2.97 | 3.74 | 4.36 | 4.78 | 1.26 | 1.47 | 1.61 |
| *c*-Bi | 3.11 | 3.48 | 4.54 | 4.62 | 1.12 | 1.46 | 1.49 |
| *a*-Bi | 3.19 | 3.75 | 4.22 | 4.76 | 1.18 | 1.33 | 1.49 |

These findings suggest that semi-metals do not follow a single structural template but rather exist in a state of "structural crossover." By identifying these transitional signatures, we complete the empirical map of non-crystalline solids, providing the necessary data to demonstrate how our renormalization approach can collapse these diverse radial and angular distributions into a single, unified framework, a Master curve.

## CONCLUSIONS

In our simulations, two interesting features appear: the existence of two-well defined families (classes), that of the amorphous semiconductors, and the one associated with the amorphous metals. But interestingly enough, despite its similarities in the characteristics that identify them as another family, the class of the semi-metallic elements clearly indicates its transitional features in going from amorphous semiconductors to amorphous metals. Whether the presence of a Master PDF can be easily identified in semiconductors and metals, however, for semi-metals, even though it exists, it is more difficult to do so, **Figure 16 (a).** This family provides a powerful lens through which to decode the structural commonalities that unify the two extremes of the disordered state, setting the stage for a universal description of amorphous structures.

In this work, we have utilized the *undermelt-quench*[19] approach to computationally generate and characterize a diverse suite of amorphous semiconductors, semi-metals, and metallic systems. Our analysis of the Pair Distribution Functions (PDFs) and Plane-Angle Distributions (PADs) demonstrates that these functions are not merely descriptive tools, but fundamental structural "fingerprints" that allow for a rigorous classification of non-crystalline solids.

Our findings lead to the following key conclusions:

- **Family-Specific Benchmarks**: We have identified distinct topological signatures for each material class. A Master PDF has been proposed for each of the families by renormalizing the data obtained and comparing them for each family (**Figures 4 (a)**, **11 (a),** and **16 (a)**). Semiconductors are defined by a "zero-minimum" between the first and second coordination shells, reflecting a rigid structural vacuum and tetrahedral coordination; metals are defined by the bimodal elephant peak and close-packing polyhedral motifs; and semi-metals by a structural crossover defined by the collapse of the inter-shell exclusion zone and the lack of close-packing motifs.

- **The Power of the PAD**: While the PDFs provide radial information, the PADs are essential for resolving the local symmetry. The shift from 109.5° (tetrahedral) to 60° (close-packing) and the appearance of 90° (orthogonal) peaks in semi-metals provide the necessary bridge to characterize the transition from covalent to metallic bonding, **Figures 4 (b)**, **11 (b),** and **16 (b)**.

- **A Unified Structural Framework**: The consistent commonalities observed within each family suggest that amorphous materials are not random, but follow invariant geometrical rules. After the renormalization, the emergence of universal scaling ratios ($r_n/r_1$), such as the $\sqrt{8/3}$ semiconductor benchmark (Second Neighbors, **Figure 4 (a), 11 (a),** and **16 (a)**) and the 1:$\sqrt{2}$:$\sqrt{3}$:2 metallic progression indicates that it is indeed possible to describe these materials through an integrated and unique mathematical fashion.

- **Methodological Validation**: The clarity of the structural features (such as the zero-minimum at 2.93 Å for a-Ge, for example) validates the *undermelt-quench* approach as a superior method for producing high-fidelity, large-scale supercells that accurately capture the amorphous state.

Ultimately, we propose that the detailed study of PDF and PAD commonalities is the most effective path toward mapping the structure of disordered systems and establishing similarities and differences among the material families. By establishing these structural boundaries, we lay the groundwork for a **renormalization approach** that can unify the diverse worlds of covalent networks, frustrated semi-metals, and metallic glasses into a single Master structural theory of amorphous materials.

## ACKNOWLEDGMENTS

I.R. and D.H.-R. thank SECIHTI and DGAPA-UNAM (POSDOC) for their respective postdoctoral fellowships. F.B.Q., S.C., and R.S.V.P. acknowledge SECIHTI for their graduate fellowships. A.A.V., R.M.V., and A.V. acknowledge DGAPA-UNAM (PAPIIT) for continued financial support under Grants No. IN118223 and IN119226. This research was also supported by SECIHTI under Grant No. CBF-2025-G-886. Computational resources were partially provided by the Supercomputing Center of DGTIC-UNAM through the project LANCAD-UNAM-DGTIC131. The authors are grateful to M. en C. Teresa Vázquez and Lic. Oralia Jiménez for their assistance with information retrieval and to Dr. Alejandro Pompa for his technical support and maintenance of the computing site at IIM-UNAM.

## AUTHOR CONTRIBUTIONS

This research was conceived and designed by I.R. and A.A.V., with input from R.M.V., A.V., and D.H.-R. Simulations were performed by I.R., D.H-R., F.B.Q., S.C., and R.S.V.P. All authors contributed to the discussion and analysis of the results. I.R. wrote the initial draft of the manuscript, which was subsequently enriched and approved by all co-authors for publication. The final revision was done by A.A.V.

## COMPETING INTERESTS STATEMENT

The authors declare no conflict of interest in this work.

## DATA AVAILABILITY STATEMENT

The datasets used and analyzed during the current study are available from the corresponding author on reasonable request.